\documentclass[twocolumn,british,authoryear,superscriptaddress,a4paper]{revtex4}
\usepackage[T1]{fontenc}
\usepackage[latin9]{inputenc}
\usepackage{color}
\usepackage{babel}
\usepackage{array}
\usepackage{textcomp}
\usepackage{amsmath}
\usepackage{graphicx}
\usepackage[unicode=true,
 bookmarks=false,
 breaklinks=false,pdfborder={0 0 1},backref=false,colorlinks=true]
 {hyperref}
\hypersetup{
 hidelinks,hidelinks,citecolor=blue,hidelinks,citecolor=blue}

\makeatletter

\providecommand{\tabularnewline}{\\}

\@ifundefined{textcolor}{}
{%
 \definecolor{BLACK}{gray}{0}
 \definecolor{WHITE}{gray}{1}
 \definecolor{RED}{rgb}{1,0,0}
 \definecolor{GREEN}{rgb}{0,1,0}
 \definecolor{BLUE}{rgb}{0,0,1}
 \definecolor{CYAN}{cmyk}{1,0,0,0}
 \definecolor{MAGENTA}{cmyk}{0,1,0,0}
 \definecolor{YELLOW}{cmyk}{0,0,1,0}
}

\makeatother

\begin{document}
\title{iCorr : Complex correlation method to detect origin of replication
in prokaryotic and eukaryotic genomes}

\author{Shubham Kundal}
\email{shubhamkundal97@gmail.com}

\affiliation{Department of Electrical Engineering, Indian Institute of Technology
(IIT) Delhi, Hauz Khas, New Delhi - 110016, India.}

\author{Raunak Lohiya}
\email{rklohiya1996@gmail.com}

\affiliation{Department of Mathematics, Indian Institute of Technology (IIT) Delhi,
Hauz Khas, New Delhi - 110016, India.}

\author{Kushal Shah}
\email{kkshah@ee.iitd.ac.in}

\affiliation{Department of Electrical Engineering, Indian Institute of Technology
(IIT) Delhi, Hauz Khas, New Delhi - 110016, India.}
\begin{abstract}
Computational prediction of origin of replication (ORI) has been of great interest in bioinformatics and several methods including GC Skew, Z curve, auto-correlation etc. have been explored in the past. In this paper, we have extended the auto-correlation method to predict ORI location with much higher resolution for prokaryotes. The proposed complex correlation method (iCorr) converts the genome sequence into a sequence of complex numbers by mapping the nucleotides to \{+1,-1,+i,-i\} instead of \{+1,-1\} used in the auto-correlation method (here, 'i' is square root of -1). Thus, the iCorr method uses information about the positions of all the four nucleotides unlike the earlier auto-correlation method which uses the positional information of only one nucleotide. Also, this earlier method required visual inspection of the obtained graphs to identify the location of origin of replication. The proposed iCorr method does away with this need and is able to identify the origin location simply by picking the peak in the iCorr graph. The iCorr method also works for a much smaller segment size compared to the earlier auto-correlation method, which can be very helpful in experimental validation of the computational predictions. We have also developed a variant of the iCorr method to predict ORI location in eukaryotes and have tested it with the experimentally known origin locations of \emph{S. cerevisiae} with an average accuracy of 71.76\%. \\
\end{abstract}

\keywords{Bioinformatics; Sequence analysis; Signal processing; Information theory.}
\maketitle

\section{Introduction}

DNA replication is a complex biological process by which the genome/chromosomes
of an organism creates a copy of itself during cell division. The
segment of DNA sequence where the process of replication initiates
on a chromosome, plasmid or virus is called origin of replication
(ORI). The ability to computationally predict ORI location is important to understand the statistical
features in DNA sequence. It could also provide information to development
of new drugs for treatment of diseases \citep{McFadden1999,Ram2007,Soldati1999}.

Prokaryotic organisms are usually found to have single origin of replication
from where two replication forks transmit in contrary directions \citep{Marians1992,Mott2007,Rocha1999}.
More evolved organisms are found to contain multiple sites from which
replication initiates and this helps to speed up the process \citep{Kelman2004,Nasheuer2002}.
Experimental detection of ORI locations is very challenging and so
far has been completed only for a very few archaea, eubacteria and
eukaryotic genomes \citep{Sernova2008}. Here computational prediction
can play a significant role by considerably reducing the search space
which can save a large amount of experimental time and effort. Computational
prediction of ORI rests on the general hypothesis that the origin
location and its flanking regions have different statistical properties
as compared to rest of the genome. Motivation for this hypothesis
comes from the fact the replication process of the leading and lagging
strands takes place through a slightly different set of proteins which
can leave certain statistical signatures at the origin location \citep{Lobry1996a,Lobry1996b}.

Different computational methods have been developed to predict origin
of replication in DNA sequence including GC-skew \citep{Lobry1996a,Lobry1996b,Mrazek1998,Touchon2005},
Z-curve \citep{Zhang2005}, CGC Skew \citep{Grigoriev1998}, AT excursion
\citep{Chew2007}, Shannon entropy \citep{Schneider1997,Schneider2010,Shannon1948},
wavelet approach \citep{Song2003}, auto-correlation based measure
\citep{Shah2012}, correlated entropy measure \citep{Parikh2015},
GC profile \citep{Li2014} and few others. All methods use the fundamental
property of identifying differences in statistical properties in the
front and end side of replication origin to account for mutational
pressures developed in the opening and ending strands of ORI \citep{Lobry2002,Mackiewicz2004}.
In the GC-skew and auto-correlation method \citep{Shah2012}, the
entire genome is divided into overlapping segments/windows and the
value of correlation measure is calculated for each window. For bacterial
genomes, usually the window size is chosen to be around one-hundredth
of the genome size and two consecutive windows have an overlap of
four-fifths of the window size. So, only one-fifth of the genome sequence
is changed per window which helps to reduce noise produced by sharp
variations of correlation measure in adjacent windows. In the GC-skew
method, the number of G and C nucleotides is counted for each segment/window
and the GC-skew value, $\left(G-C\right)\big/\left(G+C\right)$, is
plotted against the window number. An ORI is then predicted to be
present at the location where the GC-skew value crosses the zero line.
The auto-correlation method goes a step further and uses the positional
information of the G nucleotides in each window and hence is informationally
richer than the GC-skew method. It has also been shown earlier that
the auto-correlation method is able to predict the origin location
of several more genomes as compared to the GC-skew method \citep{Parikh2015,Shah2012}.

The auto-correlation method mainly has three limitations. Firstly,
the ORI location is predicted in this method by visually inspecting
the correlation profile which creates room for human error. Secondly,
the window size required in this method is quite large. Thirdly, the
auto-correlation method uses the positional information of only the
G nucleotide. In this paper, we propose a modification of this method
(iCorr) which addresses all these limitations. The proposed complex
correlation method uses four numbers $\left\{ +1,-1,+i=\sqrt{-1},-i\right\} $
and thus is able to represent the positions of each of the four nucleotides
unlike the auto-correlation method which uses only real numbers $\left\{ +1,-1\right\} $.
In the iCorr method, there is no need for visual inspection and the
ORI region is given by either the location of the peak value (for
prokaryotes) or the points of zero-crossing (for \emph{S. cerevisiae}
and perhaps other eukaryotes). This method also requires a much smaller
window size as compared to the auto-correlation method and thus leads
to a resolution that is much higher.

We describe the iCorr method in Sec. \ref{sec:Method}, present the
results in Sec. \ref{sec:Results} and finally end with discussions
in Sec. \ref{sec:Discussion}.

\section{Complex correlation method \label{sec:Method}}

The primary computational approach for prediction of origin of replication
is to divide the entire genome into overlapping windows/segments of
equal length, and analyse each window to measure some statistical
property using information theory and signal processing techniques.
The values thus obtained are plotted against the window number. The
origin of replication is predicted to be present in the window where
a significant change is observed. This abrupt change can manifest
in different ways depending on the actual statistical property being
measured.

In the auto-correlation method (henceforth, called gCorr), the G (Guanine)
nucleotide of each segment is denoted by $\left\{ +1\right\} $ and
all other nucleotides by $\left\{ -1\right\} $. This helps in converting
the symbolic sequence to a discrete number sequence thereby making
it conducive for statistical analysis. We calculate the auto-correlation
value of this discrete sequence using the function \citep{Beauchamp1979,Cavicchi2000},

\begin{equation}
C(k)=\frac{1}{(N-k)\sigma^{2}}\sum_{j=1}^{N-k}\left(a_{j}-\mu_{a}\right)\left(a_{j+k}-\mu_{a}\right)\label{eq:C-k}
\end{equation}
where $k=1,2,3,\ldots,N$, $a_{i}\in\left\{ +1,\text{\textminus}1\right\} $
denotes the value at the $i$th position of the discrete sequence,
$N$ is the window size, $\mu_{a}=0$ and $\sigma$= 1 are the means
and standard deviation of the random variable $a_{i}$. The auto-correlation
measure, $C_{G}$, is then defined as the average of all correlation
values in Eq. \eqref{eq:C-k} \citep{Shah2012},

\begin{equation}
C_{G}=\frac{1}{N-1}\sum_{k=1}^{N-1}|C(k)|\label{eq:CG}
\end{equation}
where the subscript ``G'' refers to ``genome''. $C_{G}$ ranges
from 0 to 1 and is independent of the length of the sequence. The
value of $C_{G}$ is a good indicator of the correlation strength
between the positions of the G nucleotide. Thus, a sequence with $C_{G}=0$
corresponds to a lack of correlation and one with $C_{G}=1$ to a
highly correlated sequence.

Since a DNA sequence is made up of four bases, we can generate a string
of bits for the A (Adenine) base by assigning a value of $\left\{ +1\right\} $
to every occurrence of A and $\left\{ -1\right\} $ to all other positions
(similarly for T and C). In the above method, only the G-track is
chosen for analysis since it gives much better results as compared
to the other three discrete sequences \citep{Shah2012}. Though this
method has been found to work better than the GC-skew method, it has
an inherent limitation of assigning the same value of $\left\{ -1\right\} $
to T, A and C. Due to this, it does not capture the rich variations
produced by the four bases present in DNA sequence.

In this paper, we propose the iCorr method which extends the above
method to complex states and thereby completely eliminates the most
fundamental limitation in gCorr and other computational methods for
ORI prediction. We use 
$\{+1, -1, +i = \sqrt{-1}, -i\}$
for multi-variate classification of the four bases present in a DNA
sequence. A DNA sequence made up of AGTC base pairs can give rise
to 24 different discrete sequences using the iCorr method as opposed
to only 4 sequences provided by gCorr method. After analysing all
these possible sequences, we have developed 2 variations of the iCorr
method for prokaryotic and eukaryotic organisms.

\begin{figure*}[t]
\centering{}\includegraphics[bb=0bp 0bp 640bp 480bp,clip,width=1\textwidth,trim = {4 4 4 4}]{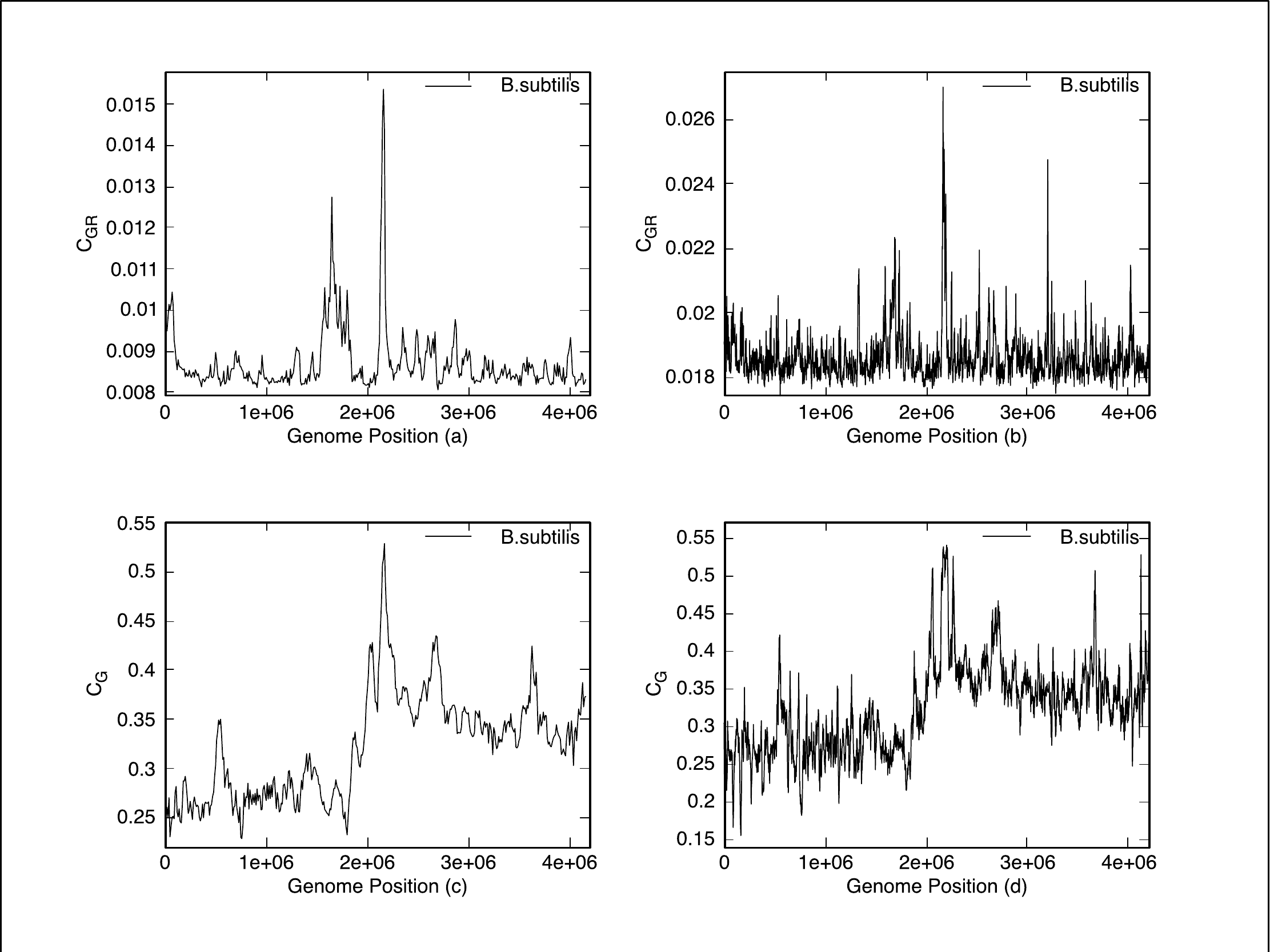}\caption{Plot of the iCorr and gCorr values for \emph{B. subtilis }(NC\_000964)\emph{.}
(a) $C_{GR}$ vs. genome position with window size = 50,000 and shift
size = 10,000. (b) $C_{GR}$ vs. genome position with window size
= 10,000 and shift size = 2,000. (c) $C_{G}$ vs. genome position
with window size = 50,000 and shift size = 10,000. (d) $C_{G}$ vs.
genome position with window size = 10,000 and shift size = 2,000.
In (a) and (b), the ORI location is given by the location of peak.
In (c) and (d), the ORI location is given by the region where $C_{G}$
undergoes an abrupt change. Clearly, the prediction of iCorr method
is much more precise and unambiguous compared to the gCorr method.\label{fig:B.subtilis} }
\end{figure*}

\begin{figure*}[t]
\begin{centering}
\includegraphics[bb=0bp 0bp 640bp 480bp,clip,width=1\textwidth,trim={4 4 4 4}]{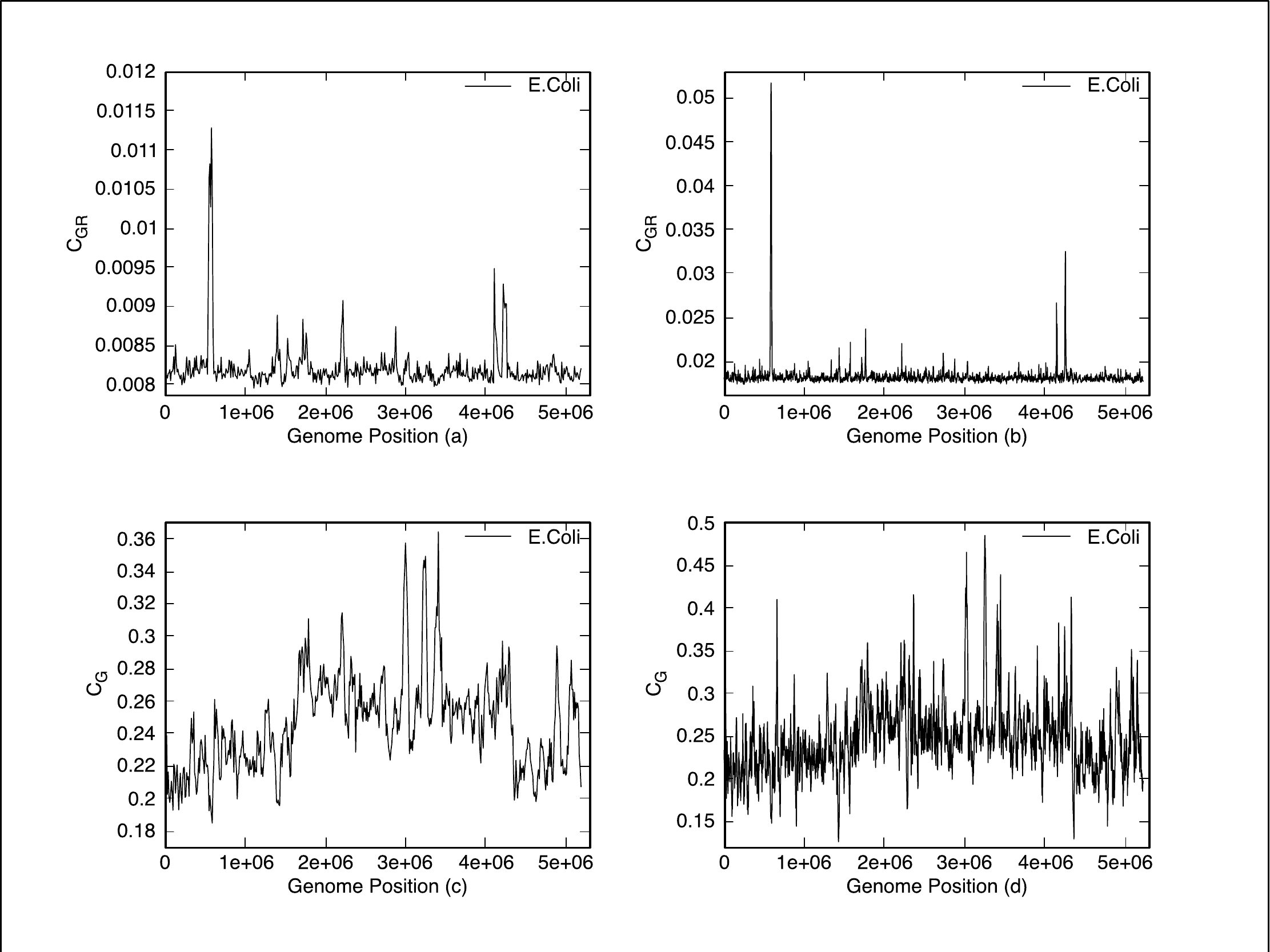} 
\par\end{centering}
\caption{Plot of the gCorr and iCorr values for \emph{E. coli} (NC\_017626)
(a) $C_{GR}$ vs. genome position with window size = 50,000 and shift
size = 10,000. (b) $C_{GR}$ vs. genome position with window size
= 10,000 and shift size = 2,000. (c) $C_{G}$ vs. genome position
with window size = 50,000 and shift size = 10,000. (d) $C_{G}$ vs.
genome position with window size = 10,000 and shift size = 2,000.
In (a) and (b), the ORI location is given by the location of peak.
In (c) and (d), the ORI location is given by the region where $C_{G}$
undergoes an abrupt change. It can be clearly seen that the prediction
capability of the gCorr method severely deteriorates as the window
size decreases, whereas the iCorr method is able to clearly predict
the ORI location for window size=10,000. \label{fig:E.coli}}
\end{figure*}

For ORI prediction in prokaryotic genomes, AGTC combination of $\left\{ +i,+1,-i,-1\right\} $
is used. We calculate the auto-correlation of this generated discrete
sequence using the same formula given in Eqs. \eqref{eq:C-k} and
\eqref{eq:CG} $\left(\text{using }\mu_{a}=0,\text{ }\sigma=1\right)$,
but now $C(k)$ comes out to be a complex number. However, the final
auto-correlation value obtained is still a real number since the RHS
of Eq. \eqref{eq:CG} uses the absolute value (or magnitude) of these
complex $C\left(k\right)$ values. This is the iCorr value for prokaryotes
(denoted by $C_{GR}$) and plotted against the window number (or genome
length). The graph produces a sharp peak as its single global property.
We propose that the genome position corresponding to this global maximum
contains the solitary origin of replication in prokaryotes.

For ORI prediction in eukaryotic bacteria, AGTC combination of $\left\{ -1,+1,-i,+i\right\} $
is used. We again calculate the auto-correlation function, $C\left(k\right)$,
of this discrete sequence using Eq. \eqref{eq:C-k}. However, for
calculating the final correlation value (denoted by $C_{GC}$) of
each segment/window, we do a sum of only the real part of $C(k)$
values,

\begin{equation}
C_{GC}=\frac{1}{N-1}\sum_{k=1}^{N-1}\Re\left\{ C(k)\right\} \label{eq:C-GC}
\end{equation}
where $\Re\left(\cdot\right)$ stands for real part of the complex
quantity within brackets. The genome positions corresponding to the
zero-crossings of the values of $C_{GC}$ are proposed to be the ORI
locations. In this way, it has some similarity to the GC-skew method
\citep{Mrazek1998}.

It is important to note here that unlike the case of prokaryotes,
we do not expect a single computational method to be able to correctly
predict ORI locations of all eukaryotic genomes due to a large amount
of variation in their statistical properties. We have tested our method
for \emph{S. cerevisiae} for which experimental results are known
and hope that this $C_{GC}$ defined above or its modifications will
be very useful in predicting the ORI locations of a wide variety of
genomes.

\section{Results\label{sec:Results}}

We have applied the method described in the previous section to 38
bacterial genomes obtained from NCBI \citep{NCBI2016} and 16 chromosomes
of one eukaryote (\emph{S. cerevisiae}) obtained from OriDB \citep{Siow2012}.
In this section, we describe the results obtained.

\subsection{ORI prediction for prokaryotes\label{Sec 3.1}}

\begin{table*}
\begin{centering}
\begin{tabular}{|>{\centering}p{1cm}|>{\centering}p{2.2cm}|>{\centering}p{2cm}|c|>{\centering}p{2cm}|c|>{\centering}p{1.8cm}|>{\centering}p{2.5cm}|}
\hline 
CHR

No.  & Experimentally Confirmed ORIs  & Window/ Shift size  & Accuracy(\%)  & Seq. Removed(\%)  & Precision(\%)  & Undetected Confirmed ORIs  & Undetected Close Confirmed ORIs\tabularnewline
\hline 
\hline 
1  & 14  & 5000/1000  & 11/14=78.57  & 35.65  & 20/47=42.55  & 3  & 0\tabularnewline
\hline 
2  & 37  & 10000/2000  & 27/37=72.97  & 32.22  & 43/90=47.77  & 10  & 3\tabularnewline
\hline 
3  & 21  & 4000/800  & 13/21=76.19  & 33.92  & 24/95=25.26  & 8  & 0\tabularnewline
\hline 
4  & 51  & 15000/300  & 37/51=72.55  & 31.13  & 63/123=51.21  & 14  & 5\tabularnewline
\hline 
5  & 22  & 5000/1000  & 14/22=63.63  & 40.52  & 19/111=17.11  & 8  & 3\tabularnewline
\hline 
6  & 17  & 3000/500  & 14/17=82.35  & 35.18  & 22/109=20.18  & 3  & 1\tabularnewline
\hline 
7  & 30  & 10000/2000  & 20/30=66.67  & 35.59  & 30/112=26.78  & 10  & 1\tabularnewline
\hline 
8  & 21  & 5000/1000  & 15/21=71.42  & 36.48  & 19/121=15.70  & 6  & 3\tabularnewline
\hline 
9  & 15  & 10000/2000  & 10/15=66.67  & 27.39  & 17/48=35.42  & 5  & 3\tabularnewline
\hline 
10  & 29  & 10000/2000  & 20/29=68.96  & 29.56  & 31/96=32.29  & 9  & 3\tabularnewline
\hline 
11  & 21  & 10000/2000  & 15/21=71.42  & 39.78  & 31/69=44.93  & 6  & 2\tabularnewline
\hline 
12  & 32  & 10000/2000  & 23/32=71.87  & 40.74  & 38/101=37.62  & 9  & 2\tabularnewline
\hline 
13  & 27  & 10000/2000  & 22/27=81.48  & 34.84  & 37/103=35.92  & 5  & 0\tabularnewline
\hline 
14  & 21  & 10000/2000  & 16/21=76.19  & 35.16  & 25/75=33.33  & 5  & 1\tabularnewline
\hline 
15  & 27  & 10000/2000  & 16/27=59.26  & 33.21  & 25/127=19.68  & 11  & 2\tabularnewline
\hline 
16  & 25  & 10000/2000  & 17/25=68  & 36.07  & 32/97=32.99  & 8  & 1\tabularnewline
\hline 
\end{tabular}
\par\end{centering}
\caption{Analysis of 16 chromosomes in \emph{S. cerevisiae}. \label{tab:Cerevisiae}}
\end{table*}

\begin{figure*}[t]
\begin{centering}
\includegraphics[clip,width=1\textwidth,trim = {4 4 4 4}]{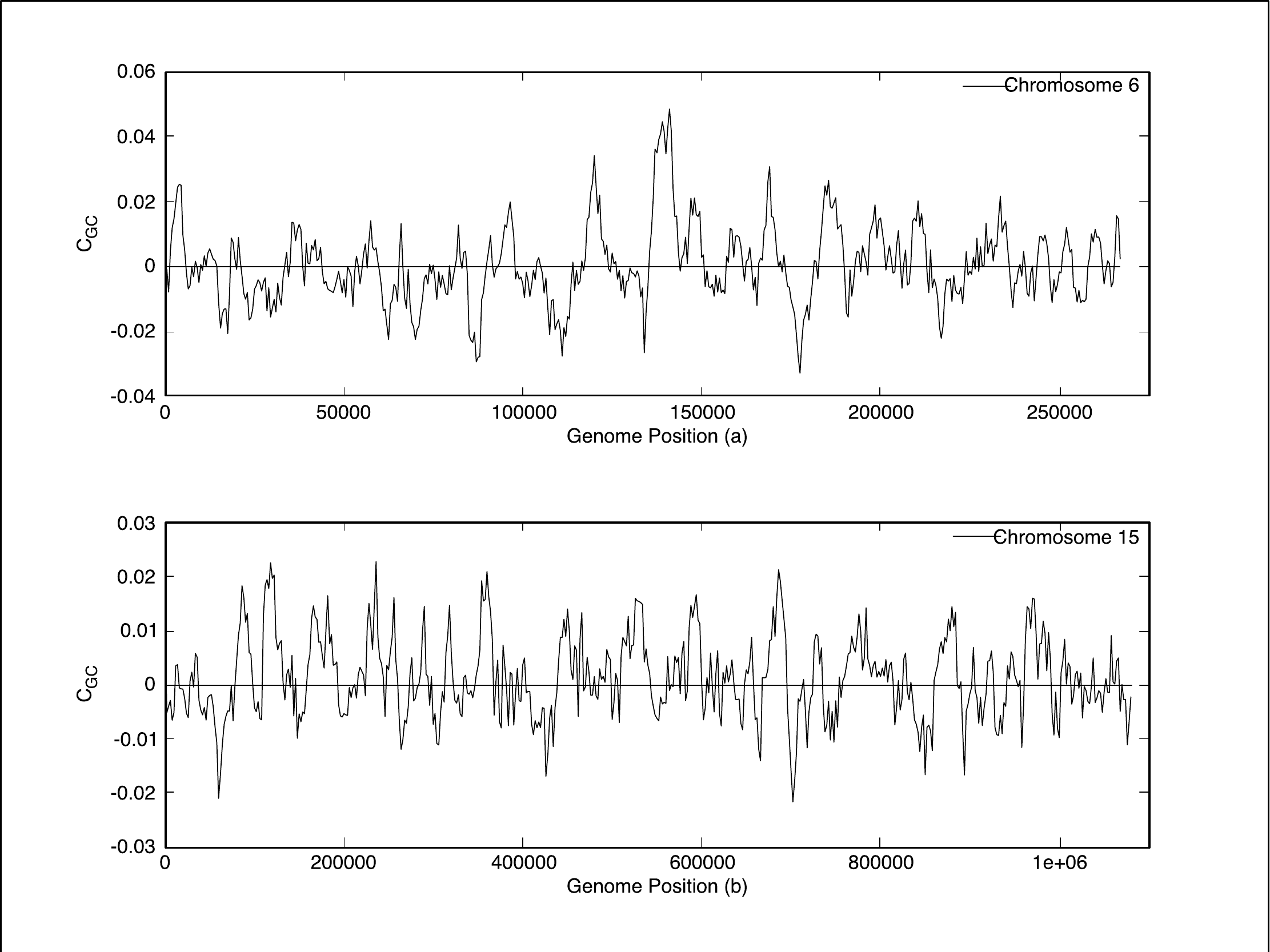} 
\par\end{centering}
\caption{Plot of iCorr vs. genome position for (a) chromosome 6 and (b) chromosome
15 in \emph{S. cerevisiae}. The points where the graph crosses the
zero-line represent possible locations of ORI. \label{fig:C-GC}}
\end{figure*}

In Figs.\ref{fig:B.subtilis} and \ref{fig:E.coli}, (a) and (c) show
graphs for iCorr and gCorr method respectively for window size of
50,000 with shift size of 10,000 while (b) and (d) show graphs for
window size of 10,000 with shift size of 2,000.

Figure\ref{fig:B.subtilis} (a) and (b) predict ORI location for \emph{B.
subtilis} using iCorr at genome positions \textcolor{black}{21,70,000}
{[}=217 (window number) \texttimes{} 10,000 (shift size){]} and 21,68,000
{[}=1,084 (window number) \texttimes{} 2,000 (shift size){]} respectively,
which are very close other. Figure \ref{fig:B.subtilis} (c) and (d)
predict ORI location using gCorr at locations of abrupt change in
the genome position ranges of 20,80,000 to 21,60,000 {[}window number
= 208-216{]} and 21,04,000 to 21,94,000 {[}window number = 1,057-1,097{]}.
Clearly, both the methods predict common genome locations but iCorr
is able to give a more precise result and reduces the genome to be
analysed for finding ORI, thereby considerably increasing the resolution.

Figure \ref{fig:E.coli} (a) and (b) predict ORI using iCorr for \emph{E.
coli} at genome positions 5,90,000 {[}=59 (window number) \texttimes{}
10,000 (shift size){]} and 5,88,000\textbf{\textcolor{red}{{} }}{[}=294
(window number) \texttimes{} 2,000 (shift size){]} respectively, which
are very close to each other. Figure \ref{fig:E.coli} (c) predicts
ORI using gCorr at locations near 15,00,000-18,00,000 genome position.
There is another abrupt change around 42,00,000-43,00,000 genome positions
which makes it very difficult to predict one ORI location using auto
correlation methods. Figure \ref{fig:E.coli} (d) uses a window size
of 10,000 and is extremely noisy and performs very poorly compared
to Fig. \ref{fig:E.coli} (c) which uses a window size of 50,000.
Therefore, the iCorr method makes a single prediction for ORI location
using both window sizes whereas gCorr fails when the window size is
small.

gCorr method predicts the presence of ORI in a genome where a sudden
transition is observed. The transition spans several windows and its
detection depends on human judgement which reduces the accuracy in
ORI prediction. In contrast, the iCorr method for prokaryotes predicts
the location by finding peak in the graph. Peak is obtained at a single
point which helps to narrow down our area of interest to a single
window. In the case of \emph{B. subtilis}, the gCorr predicts the
ORI to be present in a genome segment whose length is around 0.1 million
(see Fig.\ref{fig:B.subtilis}). In contrast, the iCorr method can
bring down the range to as low as 0.01 million genome length (20 times
higher resolution). This point is strengthened by the fact that the
obtained graph deteriorates for gCorr method as the window size is
decreased from 50,000 to 10,000 (see Fig. \ref{fig:B.subtilis} (c),
(d) and Fig. \ref{fig:E.coli} (c), (d)). The iCorr is more or less
stable and gives a fairly stable peak in the same neighbourhood even
when the window size is decreased (see Fig. \ref{fig:B.subtilis}
(a), (b) and Fig. \ref{fig:E.coli} (a), (b)). With the advantages
of peak detection and stability with window size, iCorr method is
able to predict ORI location with several times more precision than
gCorr method.

The peak by average ratio in the iCorr method was found to be in the
range $\left(1.2,\text{ }5.4\right)$ with an average of around $1.9$
in the 38 prokaryotic genomes analysed. Out of these 38 genomes, gCorr
method failed to make a clear prediction in 10 cases while iCorr faltered
in only 4 cases. However, the gCorr and iCorr method predict different
ORI locations for the same genome in many cases. In fact, only 4 instances
were found to have common prediction location out of the 38 genomes
covered. Due to lack of experimental results, we could not verify
our predictions to check which of these two methods is correct. It
is also possible that many of these genomes have multiple ORI with
different statistical properties and hence are captured by different
methods.

\subsection{ORI prediction for \emph{S. cerevisiae}\label{Sec 3.2}}

Compared to prokaryotic genomes, the computational prediction of ORI
in eukaryotic genomes has been considerably much more challenging
due to the rich and complex structure of DNA with multiple ORI being
present in a single chromosome. And an added disadvantage is that
experimentally verified ORI locations are available for only a few
eukaryotes like \emph{S. cerevisiae} and \emph{S. pombe}. The predictions
made by using gCorr and the peak detection iCorr method described
in Sec. \ref{sec:Method} do not match well with the experimental
data of these two organisms. So, we have proposed a slightly modified
version of peak detection method for this purpose and call it the
zero crossing iCorr method. The predictions of this zero crossing
method match reasonably well with the known ORI locations of \emph{S.
cerevisiae.} As described in Sec. \ref{sec:Method}, the zero crossing
iCorr method uses genome locations of occurrence of zeros instead
of the peak locations to predict multiple ORIs. Figure \ref{fig:C-GC}
shows the graph of $C_{GC}$ vs. genome location for chromosome 6
and 15 of S. \emph{cerevisiae.}

We have used different window and sub-window sizes for analysing chromosomes
to obtain optimum results (see Table \ref{tab:Cerevisiae}). The combination
of AGTC used in ORI prediction for \emph{S. cerevisiae} is $\left\{ -1,+1,-i,+i\right\} $
which is different from the combination used in ORI prediction of
bacteria. While using sliding window technique, ratio of 5:1 is maintained
between window and shift size (only chromosome 6 has a ratio of 6:1).
Table \ref{tab:Cerevisiae} summarises the data for the 16 chromosomes
analysed. The yeast chromosome sequences and data for their ORI locations
was obtained from OriDB \citep{Siow2012}.

Below is the explanation to various terms used in Table \ref{tab:Cerevisiae}: 
\begin{itemize}
\item Total Confirmed ORI : Total number of experimentally confirmed ORI
found in a chromosome as per the OriDB database. 
\item Window/ Shift Size : In the sliding window technique, window size
is the total size of each window/segment into which the genome is
divided (prediction region) and shift size is the step size, i.e.,
the amount of shift to obtain next window. 
\item Accuracy : Percentage of experimentally confirmed ORIs (as per OriDB)
which are detected by zero-crossing iCorr method. This parameter is
basically the hit rate (ratio of computationally detected and experimentally
confirmed ORI). 
\item Sequence Removed : Percentage of sequence which should not contain
ORI as per zero-crossing iCorr method. The sum of total length of
all such genomic sequence involved in the calculation of window number
where the real part of correlation measure changes sign divided by
total length of the sequence determines the parameter, ``sequence
removed''. 
\item Precision : The method predicts ORI whenever real part of ORI changes
sign. This sometimes leads to cases of false prediction, i.e. cases
where the method predicts ORI even if no confirmed ORI has been detected.
Precision is obtained by dividing the number of zero crossings which
actually contain an ORI by the total number of zero crossings. Also,
two zero-crossings can sometimes correspond to the same ORI location
since our windows/segments have been chosen to be overlapping or sometimes
an ORI location can have an overlap with two non-overlapping windows.
For the purpose of calculating precision, we count each of these zero-crossings
separately even if they point to the same ORI location.
\item Undetected Confirmed ORI : The number of confirmed ORIs at which there
was no zero crossing of $C_{GC}$. 
\item Undetected Close Confirmed ORI : Out of the undetected confirmed ORIs,
there are some ORIs which are very close to the genome position where
real part of correlation measure changes sign. The number of such
ORIs lying in the closest forward sub-window from the current window
(prediction region) are marked in this column. 
\end{itemize}
As shown in Table \ref{tab:Cerevisiae}, the accuracy for all the
16 chromosomes were in the range from 59\% to 83\% with an average
of 71.76\%. The method removed 27\% to 41\% sequence in various chromosomes
with an average of 35\%. The precision of our prediction lies in the
range from 15\% to 51\% with an average of 32.42\%. We believe this
to be a good beginning in this relatively challenging area of eukaryotic
ORI analysis, specially considering the statistical inference due
to the multiple ORI locations. 

\section{Discussion\label{sec:Discussion}}

In the past, several methods have been developed to predict ORI location
for prokaryotes but most of them utilised only a limited amount of
information present in the DNA sequence. The GC skew method \citep{Mrazek1998}
considered frequency counts of G and C nucleotides as the sole means
to predict ORI location and neglected the importance of positioning
of each base in a DNA sequence. The auto-correlation based gCorr method
was developed to remove this inherent flaw of GC skew method by considering
relative base positions of the G nucleotide. However, this method
was unable to differentiate between A, C and T nucleotides. In an
attempt to fully discover the rich variety of bases present in a sequence,
we have extended the basic gCorr method to complex states. The iCorr
method presented in this paper takes into consideration the relative
base positioning of all the four nucleotides. This method has been
found to significantly improve the resolution of ORI prediction of
prokaryotes and has also been able to predict the ORI locations of
\emph{S. cerevisiae} to a good extent. We also tried to examine the
predictions of the iCorr method for another yeast species, \emph{S.
pombe}, but the number of dubious and likely ORI positions covered
more than 90\% of the total detected ORIs. We hope that we will be
able to significantly validate and refine our methods as more experimental
data becomes available in the future.

Similar to all the previously existing computational methods, iCorr
only suggests the ORI location and does not guarantee existence of
ORI. With the advantages of pin-point peak detection and utilisation
of rich structure present in DNA, the iCorr method is a significant
progress in ORI prediction for prokaryotes. Here it is important to
note that the predictions made by these computational methods are
significantly dependent on the choice of window/segment size into
which the genome is divided for statistical analysis. If the window
size is taken to be too large, then the meaningfulness of the predictions
obviously goes down. And if the window size is taken to be too small,
the graphs can be very noise and lead to decrease in accuracy and
precision. For example, in case of chromosomes 8, 9 and 11 of \emph{S.
cerevisiae}, we applied zero-crossing method with window/shift size
of 3000/600, 2000/400 and 3000/600 respectively (here, 2000/400 means
that the window size is 2000 and shift size is 400) and precision
values dropped to 11\%, 7\% and 13\% respectively as compared to the
values reported in Table \ref{tab:Cerevisiae}.

In the iCorr method for \emph{S. cerevisiae}, only the zero-crossings
of the real part of the correlation measure given by Eq. \eqref{eq:C-GC}
have been used to predict the ORI locations. It has been observed
that the imaginary part of correlation measure remains positive for
$>99\%$ of the windows/segments and changes sign at only few isolated
contiguous points. This implies that, in some sense, we are observing
phase change of the complex correlation values to predict the location
of ORI. One interesting observation that we found was the prediction
of ORI by gCorr method always yields ORI location around the half-way
mark of the genome length. On the other hand, the predictions of ORI
by iCorr method doesn't follow any such pattern. This could be an
interesting problem to study in the future and might shed light on
the underlying statistical properties of genome sequences.

\section*{Author contributions}

SK and RL have contributed equally to this work and have carried out
the computational work and analysis. KS has designed the research
problem and analysed the results. All authors have participated in
the article preparation and approved the final article.

\end{document}